\documentclass[12pt]{article}
\usepackage{amsmath,amsfonts,amscd}

\newcommand{\hb}{\hat{\beta}}

\newcommand{\End}{\text{End}}
\newcommand{\Hom}{\text{Hom}}
\newcommand{\uqgh}{U_q(\hat{g})}

\newcommand{\one}{\mathbf{1}}
\newcommand{\id}{\text{id}}
\newcommand{\Qb}{\overline{Q}}
\newcommand{\mub}{\bar{\mu}}
\newcommand{\nub}{\bar{\nu}}
\newcommand{\sa}{{\mathcal{B}}}
\newcommand{\ab}{\alpha}

\newcommand{\ur}{{\mathcal R}}

\numberwithin{equation}{section}

\advance\textheight by \topskip
\begin{document}

\baselineskip 16pt
\parindent 10pt
\parskip 6pt


\begin{flushright}
hep-th/0112023\\ December 2001\\[3mm]
\end{flushright}
\vspace{1cm}
\begin{center}
{\Large {\bf Quantum group symmetry \\[0.05in]in sine-Gordon and
affine Toda field theories\\
[0.1in]on the half-line}}\\ \vspace{1cm} {\large G. W. Delius and
N. J. MacKay}
\\
\vspace{3mm} {\em Department of Mathematics,\\ University of York,
\\York YO10 5DD, U.K.\footnote{emails: {\tt gwd2, nm15 @york.ac.uk} }}
\end{center}

\begin{abstract}
\noindent We consider the sine-Gordon and affine Toda field
theories on the half-line with classically integrable boundary
conditions, and show that in the quantum theory a remnant survives
of the bulk quantized affine algebra symmetry generated by
non-local charges. The paper also develops a general framework for
obtaining solutions of the reflection equation by solving an
intertwining property for representations of certain coideal
subalgebras of $\uqgh$.
\end{abstract}

\section{Introduction\label{sectint}}

Affine Toda field theories are integrable relativistic quantum
field theories in 1+1 dimensions with a rich spectrum of solitons.
There is one Toda theory for each affine Kac-Moody algebra
$\hat{g}$ and the sine-Gordon model is the simplest example, for
$\hat{g}=\widehat{sl(2)}=a_1^{(1)}$.
Each of these models possesses a quantum group symmetry, generated
by non-local charges. Conservation of these charges determines the
$S$-matrices up to an overall scalar factor, as will be recalled
in section \ref{sectsqgi}.

The presence of a boundary breaks this quantum group symmetry, and
also destroys classical integrability unless the boundary
conditions are very carefully chosen. The purpose of this paper is
to show that, with such classically integrable boundary
conditions, a remnant of the quantum group symmetry nevertheless
survives. This residual symmetry is just as powerful for
determining the reflection matrices (boundary $S$-matrices) and
boundary states as the original symmetry was for determining the
particle spectrum and $S$-matrices in the bulk.

The main results of the paper are the simple expressions
\eqref{qp} and \eqref{qm} for the non-local symmetry charges in
the sine-Gordon model on the half line and their generalisations
\eqref{tcc} to $a_n^{(1)}$ affine Toda theories. The sine-Gordon
charges were first written down by Mezincescu and Nepomechie
\cite{Mez98}.

We derive the non-local symmetry charges in two very different
ways. The approach in sections \ref{sectsgm} and \ref{sectatt}
treats the boundary condition as a perturbation of the Neumann
boundary condition and employs boundary conformal perturbation
theory \cite{Gho94}. These sections thus generalize the approach
of \cite{Ber91} which derives the non-local charges on the whole
line. The second approach is purely algebraic and is introduced in
section \ref{sectcsa} and applied in section \ref{sectccrm}. While
the calculations in boundary perturbation theory presuppose a
knowledge of \cite{Ber91}, the algebraic techniques described in
section \ref{sectalg} are basic and of wider significance.

We confirm the correctness of our symmetry charges by using them
to rederive the sine-Gordon soliton reflection matrix \cite{Gho94}
in section \ref{sectrmdqgs} and the reflection matrices for the
vector solitons in $a_n^{(1)}$ affine Toda theories \cite{Gan99b}
in section \ref{sectrm}. We obtain the reflection matrices by
solving linear equations which are a consequence of the symmetry.
This is much simpler than the original approaches that relied on
solving the reflection equation \cite{Che84} (boundary Yang-Baxter
equation), which is a non-linear equation.


\section{Algebraic techniques\label{sectalg}}

\subsection{Soliton S-matrices as quantum group intertwiners\label{sectsqgi}}


This section is a review of the role of the quantum group symmetry
in determining the soliton S-matrices and serves as preparation
for the following section about reflection matrices.

The affine Toda field theory associated to the affine Lie algebra
$\hat{g}$ is known \cite{Ber91,Fel92} to be symmetric under the
quantum affine algebra $\uqgh$ with zero center. The solitons in
affine Toda field theory are arranged in various multiplets. Let
$V^\mu_\theta$ be the space spanned by the solitons in multiplet
$\mu$ with rapidity $\theta$. Each such space carries a
representation $\pi^\mu_\theta: \uqgh\rightarrow
\End(V^\mu_\theta)$. The asymptotic two-soliton states span the
tensor product spaces $V^\mu_\theta\otimes V^\nu_{\theta'}$ which
also carry representations of $\uqgh$ built using the coproduct.
An incoming two-soliton state in $V^\mu_\theta\otimes
V^\nu_{\theta'}$ with $\theta>\theta'$ will evolve during
scattering into an outgoing state in $V^\nu_{\theta'}\otimes
V^\mu_{\theta}$ with scattering amplitudes given by the S-matrix
\begin{equation}
  S^{\mu\nu}(\theta-\theta'):V^\mu_\theta\otimes
  V^\nu_{\theta'}\rightarrow V^\nu_{\theta'}\otimes
  V^\mu_{\theta}.
\end{equation}
The S-matrix has to commute with the symmetry or, in other words,
the S-matrix has to be an intertwiner of the tensor product
representations,
\begin{equation}\label{sint}
  S^{\mu\nu}(\theta-\theta')\circ
  (\pi^\mu_\theta\otimes\pi^\nu_{\theta'})
  (\Delta(Q))=
  (\pi^\nu_{\theta'}\otimes\pi^\mu_{\theta})
  (\Delta(Q))\circ
  S^{\mu\nu}(\theta-\theta')~~~\text{for all }Q\in\uqgh.
\end{equation}
Because the tensor product representations are irreducible for
generic rapidities, Schur's lemma tells us that this equation
determines the S-matrix uniquely up to an overall scalar factor.
This intertwiner can also be obtained from the universal R-matrix
${\mathcal R}$ of $\uqgh$ as
\begin{align}\label{sr}
  S^{\mu\nu}(\theta-\theta')&\propto\check{P}\circ
  R^{\mu\nu}(\theta-\theta'),&
  R^{\mu\nu}(\theta-\theta')&=
  (\pi^\mu_\theta\otimes\pi^\nu_{\theta'})({\mathcal
  R}),
\end{align}
where $\check{P}$ interchanges the tensor factors.

The scattering of three solitons is described by an intertwiner
between the tensor product representations $V^\mu_{\theta}\otimes
V^\nu_{\theta'}\otimes V^\lambda_{\theta''}$ and
$V^\lambda_{\theta''}\otimes V^\nu_{\theta'}\otimes
V^\mu_{\theta}$ with $\theta>\theta'>\theta''$. There are two ways
to build such an intertwiner from the two-soliton S-matrices:
\begin{equation}
\begin{CD}
  V^\mu_{\theta}\otimes V^\nu_{\theta'}\otimes V^\lambda_{\theta''}
  @>{S^{\mu\nu}(\theta-\theta')\,\otimes\,\id}>>
  V^\nu_{\theta'}\otimes V^\mu_{\theta}\otimes V^\lambda_{\theta''}
  @>{id\,\otimes\, S^{\mu\lambda}(\theta-\theta'')}>>
  V^\nu_{\theta'}\otimes V^\lambda_{\theta''}\otimes V^\mu_{\theta}
  \\
  @VV{\id\otimes S^{\nu\lambda}(\theta'-\theta'')}V
  @.
  @V{S^{\nu\lambda}(\theta'-\theta'')\otimes\id}VV
  \\
  V^\mu_{\theta}\otimes V^\lambda_{\theta''}\otimes V^\nu_{\theta'}
  @>{S^{\mu\lambda}(\theta-\theta'')\,\otimes\,\id}>>
  V^\lambda_{\theta''}\otimes V^\mu_{\theta}\otimes V^\nu_{\theta'}
  @>{id\,\otimes\, S^{\mu\nu}(\theta-\theta')}>>
  V^\lambda_{\theta''}\otimes V^\nu_{\theta'}\otimes V^\mu_{\theta}
\end{CD}
\end{equation}
Because the tensor product representations are irreducible for
generic rapidities, the intertwiner is unique. The above diagram
is therefore commutative (up to an overall scalar factor) -- that
is, the S-matrices satisfy the Yang-Baxter equation.

For the tensor product of two vector representations the
intertwining property \eqref{sint} was solved by Jimbo
\cite{Jim86} for all non-exceptional quantum affine algebras (both
twisted and untwisted). The intertwiners for many other tensor
products have been constructed \cite{Del94} using the tensor
product graph method. The complete sets of soliton S-matrices for
the algebras $U_q(a_n^{(1)})$ \cite{Hol92}, $U_q(c_n^{(1)})$
\cite{Gan95}, and $U_q(a^{(2)}_{2n-1})$ \cite{Gan96} have been
constructed.

\subsection{Soliton reflection matrices as intertwiners\label{sectsrmi}}

As reviewed in the previous section, quantum group symmetry can be
used to obtain the soliton S-matrices. We now want to present a
similar technique for obtaining the reflection matrices for
solitons hitting a boundary. So far the only way to obtain
reflection matrices has been to solve the reflection equation.
Because the reflection equation is a non-linear functional matrix
equation, solving it is very difficult for anything but the
simplest cases. We will instead obtain a linear equation.

We will now restrict the solitons to live on the left half line
$x\leq 0$ by imposing suitable integrable boundary conditions at
$x=0$. A soliton with positive rapidity $\theta$ will eventually
hit the boundary and be reflected into another soliton with
opposite rapidity $-\theta$. The corresponding quantum process is
described by the reflection matrix
\begin{equation}
  K^\mu(\theta): V^\mu_\theta\rightarrow V^{\mub}_{-\theta}.
\end{equation}
The multiplet $\mub$ of the reflected soliton does not necessarily
have to be the same as that of the incoming soliton, but it has to
have the same mass. It turns out \cite{Del98a} in the case of
$a_n^{(1)}$ Toda theory with the usual boundary conditions that
solitons in the r-th rank antisymmetric tensor multiplet are
converted into solitons in the $(n+1-r)$-th rank antisymmetric
tensor multiplet.

There is no $\uqgh$ intertwiner between the representations
$V^\mu_\theta$ and $V^{\mub}_{-\theta}$ -- that is, there is no
$K^\mu(\theta)$ satisfying
\begin{equation}\label{kint}
  K^\mu(\theta)\pi^\mu_\theta(Q)=\pi^{\mub}_{-\theta}(Q)K^\mu(\theta)
\end{equation}
for all $Q\in\uqgh$. This is not surprising because the boundary
should be expected to break the quantum group symmetry down to a
subalgebra $\sa$ of $\uqgh$. The intertwining property
\eqref{kint} should only hold for all $Q$ in $\sa$. The unbroken
symmetry algebra $\sa$ should be ``large enough'' so that the
intertwining condition \eqref{kint} determines the reflection
matrix uniquely up to an overall scalar factor. The subalgebra
$\sa$ must also be a left coideal of $\uqgh$ in the sense that
\begin{equation}
  \Delta(Q)\in\uqgh\otimes\sa~~~\text{ for all }Q\in\sa.
\end{equation}
This allows it to act on multi-soliton states. Later in this paper
we will construct such subalgebras $\sa$ describing the residual
quantum group symmetries of affine Toda field theories with
integrable boundary conditions.

If two solitons are incident on the boundary, they can be
reflected in two different orders. Correspondingly there are two
ways of constructing intertwiners from $V^\mu_\theta\otimes
V^\nu_{\theta'}$ to $V^{\mub}_{-\theta}\otimes
V^{\bar{\nu}}_{-\theta'}$:
\begin{equation}
\begin{CD}\label{cdre}
  V^\mu_\theta\otimes V^\nu_{\theta'}
  @>\id\,\otimes K^\nu(\theta')>>
  V^\mu_\theta\otimes V^{\bar{\nu}}_{-\theta'}
  @>S^{\mu\bar{\nu}}(\theta+\theta')>>
  V^{\bar{\nu}}_{-\theta'}\otimes V^\mu_\theta
  @>\id\,\otimes K^\mu(\theta)>>
  V^{\bar{\nu}}_{-\theta'}\otimes V^{\bar{\mu}}_{-\theta}\\
  @VVS^{\mu\nu}(\theta-\theta')V
  @.@.
  @VS^{\bar{\nu}\bar{\mu}}(\theta-\theta')VV\\
  V^\nu_{\theta'}\otimes V^\mu_{\theta}
  @>\id\,\otimes K^\mu(\theta)>>
  V^\nu_{\theta'}\otimes V^{\bar{\mu}}_{-\theta}
  @>S^{\nu\bar{\mu}}(\theta+\theta')>>
  V^{\bar{\mu}}_{-\theta}\otimes V^\nu_{\theta'}
  @>\id\,\otimes K^\nu(\theta')>>
  V^{\bar{\mu}}_{-\theta}\otimes V^{\bar{\nu}}_{-\theta'}
\end{CD}
\end{equation}
Provided the tensor product representations are irreducible as
representations of the subalgebra $\sa$ the diagram above is
commutative (up to an overall scalar factor) -- that is, the
reflection matrix automatically satisfies the reflection equation.

Due to the identification \eqref{sr} between the S-matrix and the
R-matrix the reflection equation can also be written in the form
\begin{equation}\label{rn}
  \check{P}R^{\nub\mub}(\theta-\theta')\check{P}
  \,\overset{1}{K^\mu}(\theta)\,
  R^{\mu\nub}(\theta+\theta')
  \,\overset{2}{K^\nu}(\theta')
  =
  \overset{2}{K^\nu}(\theta')\,
  \check{P}R^{\nu\mub}(\theta+\theta')\check{P}
  \,\overset{1}{K^\mu}(\theta)\,
  R^{\mu\nu}(\theta-\theta'),
\end{equation}
where we employ the standard notation $\overset{1}{A}=A\otimes
\id$, $\overset{2}{A}=\id\otimes A$. Note that there is one such
reflection equation for every pair of soliton multiplets and that
generally these reflection equations involve 4 different
$R$-matrices.

In affine Toda theory it is possible for solitons to bind to the
boundary, thereby creating multiplets of boundary bound states.
The space $V^{[\lambda]}$  spanned by the boundary bound states in
multiplet $[\lambda]$ will carry a representation
$\pi^{[\lambda]}:\sa\rightarrow\End(V^{[\lambda]})$ of the
symmetry algebra $\sa$. The reflection of solitons in multiplet
$\mu$ with rapidity $\theta$ off a boundary bound state in
multiplet $[\lambda]$ is described by a reflection matrix
$K^{\mu[\lambda]}(\theta):V^\mu_\theta\otimes
V^{[\lambda]}\rightarrow V^{\mub}_{-\theta}\otimes V^{[\lambda]}$
which is determined by the intertwining property
\begin{equation}
  K^{\mu[\lambda]}(\theta)\,(\pi^\mu_\theta\otimes
  \pi^{[\lambda]})(\Delta(Q))=(\pi^{\mub}_{-\theta}\otimes
  \pi^{[\lambda]})(\Delta(Q))\,K^{\mu[\lambda]}(\theta),~~~~
  \forall \,Q\in\sa.
\end{equation}

\subsection{Construction of symmetry algebra\label{sectcsa}}

In this paper we will use boundary conformal perturbation theory
to construct generators of the coideal subalgebras
$\sa_\epsilon\subset\uqgh$ that occur as the symmetry algebras of
affine Toda field theories on the half line, where $\epsilon$
parameterizes the boundary condition. However we also have an
alternative construction which we will describe in this section.
The construction has the disadvantage that it requires the a
priori knowledge of at least one solution of the reflection
equation but it has the advantage that it does not rely on first
order perturbation theory. We will use it in section
\ref{sectccrm} to verify the expressions for the symmetry charges
derived in section \ref{sectcccpt}.

Let us assume that for one particular representation
$V^\mu_\theta$ we know the reflection matrix
$K^\mu(\theta):V^\mu_\theta\rightarrow V^{\bar{\mu}}_{-\theta}$.
We define the corresponding $\uqgh$-valued $L$-operators
\cite{FRT} in terms of the universal $R$-matrix ${\mathcal R}$ of
$\uqgh$ \cite{Kho92},
\begin{align}
  L^\mu_\theta&=\left(\pi^\mu_\theta\otimes\id\right)
  ({\mathcal R})\in\End(V^\mu_\theta)\otimes\uqgh,\notag\\
  \bar{L}^{\bar{\mu}}_\theta&=
  \left(\pi^{\bar{\mu}}_{-\theta}\otimes\id\right)
  ({\mathcal R}^{\text{op}})\in\End(V^{\mub}_{-\theta})\otimes\uqgh.\label{l}
\end{align}
Here $\ur^{\text{op}}$ is the opposite universal R-matrix obtained
by interchanging the two tensor factors. Motivated by \cite{skl}
we construct the matrices
\begin{equation}\label{b}
  B^\mu_\theta=\bar{L}^{\bar{\mu}}_\theta\,
  (K^\mu(\theta)\otimes\,1)\,L^\mu_\theta\in
  \Hom(V^\mu_\theta,V^{\mub}_{-\theta})\otimes\uqgh.
\end{equation}
It may make things clearer to introduce matrix indices:
\begin{equation}
  (B^\mu_\theta)^\alpha{}_\beta=
  (\bar{L}^{\bar{\mu}}_\theta)^\alpha{}_\gamma
  (K^\mu(\theta))^\gamma{}_\delta(L^\mu_\theta)^\delta{}_\beta
  \in\uqgh,
\end{equation}
where we are using the usual summation convention.
We will now check that for all $\theta$ the
$(B^\mu_\theta)^\alpha{}_\beta$ are elements of a coideal
subalgebra $\sa$ which commutes with the reflection matrices.

Let us first check that any reflection matrix
$K^\nu(\theta'):V^\nu_{\theta'}\rightarrow V^{\nub}_{-\theta'}$
which satisfies the appropriate reflection equation commutes with
the action of the elements $(B^\mu_\theta)^\alpha{}_\beta$, i.e.,
that (see eq. \eqref{kint})
\begin{equation}
  K^\nu(\theta')\circ\pi^\nu_{\theta'}((B^\mu_\theta)^\alpha{}_\beta)=
  \pi^{\bar{\nu}}_{-\theta'}((B^\mu_\theta)^\alpha{}_\beta)\circ K^\nu(\theta'),
\end{equation}
or, in index-free notation,
\begin{equation}\label{inc}
  (\id\otimes K^\nu(\theta'))\circ
  (\id\otimes \pi^\nu_{\theta'})(B^\mu_\theta)=
  (\id\otimes \pi^{\bar{\nu}}_{-\theta'})(B^\mu_\theta)\circ
  (\id\otimes K^\nu(\theta')).
\end{equation}
We observe that
\begin{align}
  (\id\otimes \pi^\nu_{\theta'})(L^\mu_\theta)&=
  (\pi^\mu_\theta\otimes\pi^\nu_{\theta'})({\mathcal R})=
  R^{\mu\nu}(\theta-\theta'),\\
  (\id\otimes \pi^\nu_{\theta'})(\bar{L}^{\mub}_\theta)&=
  (\pi^{\mub}_{-\theta}\otimes\pi^\nu_{\theta'})
  ({\mathcal R}^{\text{op}})=\check{P}
  R^{\nu\mub}(\theta+\theta')\check{P},\\
  (\id\otimes \pi^{\nub}_{-\theta'})(L^\mu_\theta)&=
  (\pi^\mu_\theta\otimes\pi^{\nub}_{-\theta'})({\mathcal R})=
  R^{\mu\nub}(\theta+\theta'),\\
  (\id\otimes \pi^{\nub}_{-\theta'})(\bar{L}^{\mub}_\theta)&=
  (\pi^{\mub}_{-\theta}\otimes\pi^{\nub}_{-\theta'})
  ({\mathcal R}^{\text{op}})=\check{P}
  R^{\nub\mub}(\theta-\theta')\check{P},
\end{align}
Substituting this into \eqref{inc} and using $\check{P}R\propto S$
gives
\begin{multline}\label{ree}
  (\id\otimes K^\nu(\theta'))\circ
  S^{\nu\mub}(\theta+\theta')\circ
  (\id\otimes K^\mu(\theta))\circ
  S^{\mu\nu}(\theta-\theta')\\
  =S^{\nub\mub}(\theta-\theta')\circ
  (\id\otimes K^\mu(\theta))\circ
  S^{\mu\nub}(\theta+\theta')\circ
  (\id\otimes K^\nu(\theta')),
\end{multline}
which is just the reflection equation (compare with \eqref{cdre}).
We thus see that every solution of the reflection equation
commutes with all the generators $(B^\mu_\theta)^\alpha{}_\beta$,
and vice versa: every matrix which satisfies the intertwining
equation \eqref{inc} is also a solution of the reflection equation
\eqref{ree}.

Next we need to check the coideal property. Under the assumption
that all $(B^{\mu}_\theta)^\alpha{}_\beta$ are in $\sa$ we need to
show that $\Delta\left((B^{\mu}_\theta)^\alpha{}_\beta\right)$ is
in $\uqgh\otimes\sa$. Using that
\begin{align}
  \Delta\left((L^\mu_\theta)^\alpha{}_\beta\right)
  &=\left((\pi^\mu_\theta)^\alpha{}_\beta\otimes\Delta\right)
  ({\mathcal R})\\
  &=\left((\pi^\mu_\theta)^\alpha{}_\beta\otimes\id\otimes\id\right)
  ({\mathcal R}_{13}{\mathcal R}_{12})\\
  &=(L^\mu_\theta)^\gamma{}_\beta\otimes
  (L^\mu_\theta)^\alpha{}_\gamma,
\end{align}
and similarly
\begin{equation}
  \Delta\left((\bar{L}^{\mub}_\theta)^\alpha{}_\beta\right)
  =(\bar{L}^{\mub}_\theta)^\alpha{}_\gamma\otimes
  (\bar{L}^{\mub}_\theta)^\gamma{}_\beta,
\end{equation}
we find that
\begin{equation}
  \Delta\left((B^{\mu}_\theta)^\alpha{}_\beta\right)
  =(\bar{L}^{\mub}_\theta)^\alpha{}_\delta
  (L^\mu_\theta)^\sigma{}_\beta
  \otimes (B^{\mu}_\theta)^\delta{}_\sigma,
\end{equation}
which is indeed in $\uqgh\otimes\sa$ as required.

Note that the $B$-matrices satisfy the quadratic relations
\begin{equation}
  \check{P}R^{\nub\mub}(\theta-\theta')\check{P}\,
  \overset{1}{B^\mu_\theta}\,
  R^{\mu\nub}(\theta+\theta')\,
  \overset{2}{B^\nu_{\theta'}}
  =\overset{2}{B^\nu_{\theta'}}\,
  \check{P}R^{\nu\mub}(\theta+\theta')\check{P}\,
  \overset{1}{B^\mu_\theta}\,
  R^{\mu\nu}(\theta-\theta'),
\end{equation}
which follow from the FRT relations satisfied by the $L$-matrices
together with the reflection equation \eqref{rn}. Algebras with
relations of this form are known as reflection equation algebras
\cite{skl}. Our construction can thus be viewed as an embedding of
the reflection equation algebras into the quantized enveloping
algebras.

%

\section{The sine-Gordon model\label{sectsgm}}

\subsection{Review of non-local charges\label{sectrnlc}}

For the purpose of finding its quantum group symmetry charges one
views the sine-Gordon model as a perturbation of a free bosonic
conformal field theory by a relevant operator $\Phi^{\rm
pert}$\cite{zam}. The (Euclidean) action on the whole line
is\footnote{We use the conventions of \cite{Ber91} and denote
Euclidean light-cone coordinates as $z=(t+ix)/2$ and
$\bar{z}=(t-ix)/2$, where the Euclidean time $t$ is related to
Minkowski time $t^M$ by $t=it^M$. The derivatives are then
$\partial=\partial_t-i\partial_x,\bar{\partial}=\partial_t+i\partial_x$.
We write $d^2z=idz\,d\bar{z}=-dx\,dt/2$.} \begin{equation} S =
\frac{1}{4\pi} \int d^2z\,\partial\phi \bar{\partial}\phi
\;+\;\frac{\lambda}{2\pi} \int d^2z\, \Phi^{\rm pert}(x,t)\,,
\end{equation} with the perturbing operator \begin{equation}
\Phi^{\rm pert}(x,t)= e^{i\hb\phi(x,t)} + e^{-i\hb\phi(x,t)}\,,
\end{equation} where $\hb$ is the coupling constant\footnote{$\hb$
is related to the conventional $\beta$ by
$\hb=\beta/\sqrt{4\pi}$.}. We impose the condition
$\phi(-\infty,t)=0$.

The free boson field may be split into holomorphic and
antiholomorphic parts, $\phi=\varphi+\bar{\varphi}$, where
$\bar{\partial}\varphi=0=\partial\bar{\varphi}$, and the two-point
functions are \begin{equation}\label{tp} \langle
\varphi(z)\varphi(w)\rangle_0=-\ln(z-w),\qquad  \langle
\bar{\varphi}(\bar{z})\bar{\varphi}(\bar{w})\rangle_0=-\ln(\bar{z}-\bar{w}),
\qquad  \langle
\varphi(z)\bar{\varphi}(\bar{w})\rangle_0=0\,.\end{equation} The
set of fields in the conformal field theory consists only of those
combinations of derivatives and exponentials of the fundamental
fields $\varphi$ and $\bar{\varphi}$ which do not suffer from
logarithmic divergences and are local with respect to each other.
See \cite{Kla92} for a clear account of the free bosonic theory
and its perturbation into the sine-Gordon model.

The $U_q(\hat{sl_2})$ symmetry of the sine-Gordon model is
generated by the non-local charges \cite{Ber91}
\begin{equation}
  Q_\pm = \frac{1}{4\pi} \int_{-\infty}^\infty
  dx(J_\pm-H_\pm)\,,\qquad \bar{Q}_\pm = \frac{1}{4\pi}
  \int_{-\infty}^\infty dx ( \bar{J}_\pm-\bar{H}_\pm)\,,
\end{equation}
where
\begin{eqnarray}
  &J_\pm = : e^{\pm \frac{2i}{\hb}\varphi}:\,,
  \qquad\bar{J}_\pm = : e^{\mp
  \frac{2i}{\hb}\bar{\varphi}}:\,,&\\[0.1in]& H_\pm =\lambda
  {\hb^2\over \hb^2-2} : \exp\left( \pm i
  \left({2\over\hb}-\hb\right)\varphi \mp
  i\hb\bar{\varphi}\right):&\,,\\[0.1in]
  &\bar{H}_\pm =\lambda {\hb^2\over \hb^2-2} : \exp\left( \mp i
  \left({2\over\hb}-\hb\right)\bar{\varphi} \pm
  i\hb\varphi\right):&\,
\end{eqnarray}
together with the topological charge
\begin{equation}
  T = {\hb\over 2\pi}
 \int_{-\infty}^\infty dx\,\partial_x \phi\,.
\end{equation}
The time-independence of the charges follows from the current
conservation equations \begin{equation} \bar{\partial} J_\pm=
\partial H_\pm\,,\quad \partial\bar{J}_\pm =
\bar{\partial}\bar{H}_\pm\,,\end{equation} which were obtained in
first-order perturbation theory in \cite{Ber91}. In order to
derive the $U_q(\hat{sl_2})$ relations and coproduct, we set
$\phi(-\infty)=0$. The parameter $q$ is then related to the Toda
coupling constant $\hb$ by
\begin{equation}\label{q}
  q=\exp\left(2i\pi(1-\hb^2)/\hb^2\right).
\end{equation}

\subsection{Neumann boundary condition\label{sectnbc}}

We now want to restrict the sine-Gordon model to the half line
$x\leq 0$ by imposing the Neumann boundary condition
$\partial_x\tilde{\phi}=0$ at $x=0$. Note that to avoid confusion
we will always decorate fields in the theory on the half-line with
a tilde. Again we will consider the sine-Gordon model as a
perturbation of the free boson.

A simple way to describe the free bosonic field theory on the
half-line with Neumann boundary condition is to identify its
fields with the subset of parity-invariant fields of the theory on
the whole line. Thus for every field $\Phi(x,t)$ on the whole line
there exists a field $\tilde{\Phi}(x,t)$ on the half-line defined
by
\begin{equation}\label{hl}
\tilde{\Phi}(x,t) = \Phi(x,t)+\bar{\Phi}(-x,t)
\qquad {\rm for}\;\; x\leq 0\,,\end{equation} where
$\bar{\Phi}(-x,t)$ is the parity-transform of $\Phi(x,t)$. The
fundamental field on the half-line, $\tilde{\phi}(x,t) = \phi(x,t)
+ \phi(-x,t)$, immediately satisfies the Neumann boundary
condition. The two-point functions for its chiral components
\begin{equation} \tilde{\varphi}(x,t)= \varphi(x,t) +
\bar{\varphi}(-x,t) \qquad {\rm and} \qquad
\tilde{\bar{\varphi}}(x,t) = \bar{\varphi}(x,t) + \varphi(-x,t)
\end{equation} follow immediately from \eqref{tp}, and are \begin{eqnarray} & \langle
\tilde{\varphi}(z)  \tilde{\varphi}(w)\rangle_0 =
-2\ln(z-w)\,,\qquad \langle \tilde{\bar{\varphi}}(z)
\tilde{\bar{\varphi}}(w)\rangle_0 = -2\ln(\bar{z}-\bar{w})\,,&
\\ & \langle \tilde{\varphi}(z)
\tilde{\bar{\varphi}}(\bar{w})\rangle_0 = -2\ln(z-\bar{w})\,.&
\end{eqnarray}
Note the non-vanishing two-point function between the
holomorphic and anti-holomorphic components.

The perturbation $\int d^2z\,\Phi^{\rm pert}$ is invariant under
parity and is thus a valid perturbation of the boson on the
half-line too. Note that one does not obtain the sine-Gordon model
on the half-line by perturbing with the operator
$\exp(i\hb\tilde{\phi}(x,t))+\exp(-i\hb\tilde{\phi}(x,t))$, as one
might naively have thought\footnote{See however \cite[appendix
C]{Baj01} for a different treatment of the boundary perturbation}.

The charges  $T, Q_\pm$ and $\bar{Q}_\pm$ described in the
previous section transform under parity ${\cal P}$ as follows:
\begin{equation} {\cal P} : T \mapsto -T, \qquad Q_\pm \mapsto
\bar{Q}_\mp , \qquad \bar{Q}_\pm \mapsto Q_\mp\,.\end{equation}
The parity-invariant combinations $ \tilde{Q}_\pm = Q_\pm +
\bar{Q}_\mp$ therefore give conserved charges on the half-line,
but $T$ does not. We can express the charges $\tilde{Q}_\pm$ in
terms of currents on the half-line,
\begin{equation}\label{qce}
\tilde{Q}_\pm = {1\over 4\pi} \int_{-\infty}^0 dx\left(
\tilde{J}_\pm - \tilde{H}_\pm +\tilde{\bar{J}}_\mp -
\tilde{\bar{H}}_\mp\right),
\end{equation}
where the half-line currents are defined according to \eqref{hl},
for example $\tilde{J}_\pm(x)=J_\pm(x)+\bar{J}_\mp(-x)$. They
satisfy the conservation equations \begin{equation}
\bar{\partial}\tilde{J}_\pm =
\partial \tilde{H}_\pm \qquad {\rm and} \qquad \partial \tilde{\bar{J}}_\pm =
\bar{\partial}\tilde{\bar{H}}_\pm\,.\end{equation}

\subsection{General boundary conditions as boundary perturbations\label{sectgbcbp}}

In \cite{Skl88,Gho94} the sine-Gordon model was found to be
classically integrable with the rather more general boundary
condition
\begin{equation}\label{gbc}
  \partial_x \tilde\phi = i\hb\lambda_b \left( \epsilon_- e^{i\hb
  \tilde{\phi}(0,t)/2} -   \epsilon_+ e^{-i\hb \tilde{\phi}(0,t)/2}
  \right)\qquad{\rm at}\;\; x=0\,.
\end{equation}
We shall treat this as a boundary perturbation of the sine-Gordon
model with Neumann boundary condition:
\begin{equation}
  S_\epsilon = S_{\rm Neumann} +{\lambda_b\over 2\pi} \int dt\,
  \Phi^{\rm pert}_{\rm boundary}(t)\,,
\end{equation}
with the boundary perturbing operator
\begin{align}\label{gbp}
  \Phi^{\rm pert}_{\rm boundary}(t) & =  \epsilon_- e^{i\hb
  \tilde{\phi}(0,t)/2} + \epsilon_+ e^{-i\hb
  \tilde{\phi}(0,t)/2}\\
  &=\epsilon_- e^{i\hb \phi(0,t)} + \epsilon_+
  e^{-i\hb \phi(0,t)}\,.\notag
\end{align}

To check that $S_\epsilon$ does indeed produce the boundary
condition \eqref{gbc} we calculate the correlation functions of
$\partial_x\tilde{\phi}(0,t)$ in first-order boundary conformal
perturbation theory \cite{Gho94,Pen95}:
\begin{align}
  \langle \partial_x\tilde{\phi}(0,t)\cdots\rangle
  &=\lim_{x\rightarrow 0^-}\langle \partial_x\tilde{\phi}(x,t)
  e^{-S_{\text{boundary}}}\cdots\rangle_{\text{N}}\\
  &=\langle \partial_x\tilde{\phi}(0,t)\cdots\rangle_{\text{N}}
  -\frac{\lambda_b}{2\pi}\int dt'
  \lim_{x\rightarrow 0^-}\langle \partial_x\tilde{\phi}(x,t)
  \Phi^{\rm pert}_{\rm boundary}(t')\cdots\rangle_{\text{N}}
  +{\mathcal O}(\lambda_b^2),\nonumber
\end{align}
where $\langle \cdots\rangle_{\text N}$ denotes the correlation
function with Neumann boundary condition. The first term on the
right hand side vanishes of course. To evaluate the second we need
the operator product expansions
\begin{align*}
  \partial\tilde{\varphi}(x,t)\Phi^{\rm pert}_{\rm boundary}(t')&=
  \frac{-2i\hb}{t-t'+ix}\left(
  \epsilon_- e^{i\hb\tilde{\phi}(0,t)/2}-
  \epsilon_+e^{-i\hb\tilde{\phi}(0,t)/2}\right)
  +\text{regular terms},\\
  \overline{\partial}\tilde{\bar{\varphi}}(x,t)
  \Phi^{\rm pert}_{\rm boundary}(t')&=
  \frac{-2i\hb}{t-t'-ix}\left(
  \epsilon_- e^{i\hb\tilde{\phi}(0,t)/2}-
  \epsilon_+e^{-i\hb\tilde{\phi}(0,t)/2}\right)
  +\text{regular terms},
\end{align*}
and thus, using that $\partial_x\phi=\frac
i2(\partial\tilde{\varphi}-\bar{\partial}{\tilde{\bar{\varphi}}}+{\mathcal
O}(\lambda_b)$,
\begin{equation}
  \partial_x\phi(x,t)\Phi^{\rm pert}_{\rm boundary}(t')=\hb
  \left(\frac{1}{t-t'+ix}-\frac{1}{t-t'-ix}\right)\left(
  \epsilon_- e^{i\hb\tilde{\phi}(0,t)/2}-
  \epsilon_+e^{-i\hb\tilde{\phi}(0,t)/2}\right)
  +\dots.
\end{equation}
We can now use the identity
\begin{equation}
  \lim_{x\rightarrow 0^-}
  \left(\frac{1}{(t-t'+ix)^n}-\frac{1}{(t-t'-ix)^n}\right)
  =\frac{2\pi i}{(n-1)!}\partial_{t'}^{n-1}\delta(t-t'),
\end{equation}
to find
\begin{equation}
  \langle \partial_x\tilde{\phi}(0,t)\cdots\rangle=
  \langle i\hb\lambda_b\left(
  \epsilon_- e^{i\hb\tilde{\phi}(0,t)/2}-
  \epsilon_+e^{-i\hb\tilde{\phi}(0,t)/2}\right)\rangle,
\end{equation}
in agreement with the boundary condition \eqref{gbc}.

\subsection{Quantum group charges for general boundary condition\label{sectqgcgbc}}

For the general boundary conditions \eqref{gbc} (that is, in the
presence of the boundary perturbation \eqref{gbp}) the charges
$\tilde{Q}_\pm$ in \eqref{qce} will no longer be conserved. We
calculate the time-dependence of the charges:
\begin{eqnarray*}
\partial_t \tilde{Q}_\pm & = & {1\over 4\pi} \int_{-\infty}^0
dx\,\partial_t\left( \tilde{J}_\pm - \tilde{H}_\pm
+\tilde{\bar{J}}_\mp - \tilde{\bar{H}}_\mp\right) \\ & = &
-{i\over 4\pi} \int_{-\infty}^0 dx\,\partial_x\left( \tilde{J}_\pm
+ \tilde{H}_\pm -\tilde{\bar{J}}_\mp - \tilde{\bar{H}}_\mp\right)
\\
& = &-{i\over 4\pi}\left( \tilde{J}_\pm(0,t) + \tilde{H}_\pm(0,t)
-\tilde{\bar{J}}_\mp(0,t) - \tilde{\bar{H}}_\mp(0,t)\right)\,.
\end{eqnarray*}
For the Neumann boundary condition, $J_\pm(0,t) =
\bar{J}_\mp(0,t)$ and $H_\pm(0,t) = \bar{H}_\mp(0,t)$ and thus
$\partial_t\tilde{Q}_\pm=0$. But in general we will obtain in
first order perturbation theory
\begin{align}
  \langle \partial_t \tilde{Q}_+(0,t)\cdots\rangle
  &=-\frac{\lambda_b}{2\pi}\int dt'\lim_{x\rightarrow 0^-}
  \langle\partial_t\tilde{Q}_+(x,t)\Phi^{\rm pert}_{\rm boundary}(t')
  \dots\rangle_N+\dots\\
  &=-\frac{\lambda_b}{2\pi}\int dt'\lim_{x\rightarrow 0^-}
  \langle\frac{-i}{4\pi}
  \left(\tilde{J}_+(x,t)-\tilde{\bar{J}}_-(x,t)\right)
  \Phi^{\rm pert}_{\rm boundary}(t')
  \dots\rangle_N+\dots.
\end{align}
Because $\tilde{H}_+$ and $\tilde{\bar{H}}_-$ are themselves
already of order $\lambda$ they do not contribute to first order.
The necessary operator product expansions are
\begin{align*}
  \tilde{J}_+(x,t)\Phi^{\rm pert}_{\rm boundary}(t')&=
  \epsilon_+\frac{1}{(t-t'+ix)^2}
  :\left(e^{\frac{2i}{\hb}\varphi(x,t)}
  +e^{\frac{2i}{\hb}\bar{\varphi}(-x,t)}
  \right)e^{-i\hb\phi(0,t)}:
  +\dots,\\
  \tilde{\bar{J}}_-(x,t)\Phi^{\rm pert}_{\rm boundary}(t')&=
  \epsilon_+\frac{1}{(t-t'-ix)^2}
  :\left(e^{\frac{2i}{\hb}\varphi(x,t)}
  +e^{\frac{2i}{\hb}\bar{\varphi}(-x,t)}
  \right)e^{-i\hb\phi(0,t)}:
  +\cdots.\\
\end{align*}
Using now that at the boundary $\bar{\varphi}=\varphi=\phi/2$ up
to order $\lambda$ terms, we obtain
\begin{align*}
  \langle \partial_t \tilde{Q}_+(0,t)\cdots\rangle
  &=-\frac{\lambda_b\epsilon_+}{2\pi}\frac{1}{2\pi i}
  \int dt'\lim_{x\rightarrow 0^-}
  \left(\frac{1}{(t-t'+ix)^2}-\frac{1}{(t-t'-ix)^2}\right)\\
  &~~~~~~~~~~~~~
  :e^{\frac{i}{\hb}\phi(x,t)}e^{-i\hb\phi(0,t')}:+\cdots\\
  &=\frac{\lambda_b\epsilon_+}{2\pi}\frac{\hb^2}{\hb^2-1}
  \partial_t e^{i\left(\frac{1}{\hb}-\hb\right)\phi(0,t)}+\cdots\\
  &=\frac{\lambda_b\epsilon_+}{2\pi}\frac{\hb^2}{\hb^2-1}
  \partial_t q^{T}+\cdots,
\end{align*}
where we used that $q$ has the value given in \eqref{q}. It
follows that the charge
\begin{equation}\label{qp}
  \hat{Q}_+=Q_++\bar{Q}_-+\hat{\epsilon}_+q^{T}~~~\text{with }~~
  \hat{\epsilon}_+=\frac{\lambda_b\epsilon_+}{2\pi}\frac{\hb^2}{1-\hb^2}
\end{equation}
is conserved to first order in perturbation theory. By similar
calculations we obtain a second conserved charge
\begin{equation}\label{qm}
  \hat{Q}_-=Q_-+\bar{Q}_++\hat{\epsilon}_-q^{-T}~~~\text{with }~~
  \hat{\epsilon}_-=\frac{\lambda_b\epsilon_-}{2\pi}\frac{\hb^2}{1-\hb^2}.
\end{equation}
These charges were first written down in \cite{Mez98}. They
generate a coideal subalgebra of $\uqgh$ because
\begin{equation}
  \Delta(\hat{Q}_\pm)=(Q_\pm+\bar{Q}_\mp)\otimes 1+q^{\pm
  T}\otimes\hat{Q}_\pm.
\end{equation}

\subsection{Reflection matrices derived from quantum group
symmetry\label{sectrmdqgs}}

We will now use our knowledge of the conserved charges
$\hat{Q}_\pm$ to derive the soliton reflection matrix, up to an
overall factor.

The sine-Gordon model only has a single two-dimensional soliton
multiplet, spanned by the soliton and anti-soliton states
$|A_\pm(\theta)\rangle$. The soliton reflection matrix describes
what happens to a soliton during reflection off the boundary: a
soliton of type $\alpha$ with rapidity $\theta$ is converted into
a combination of soliton types $\beta$ with opposite rapidity
$-\theta$ with probability amplitudes $K^\beta{}_\alpha$,
\begin{equation}
  K:|A_\alpha(\theta)\rangle\mapsto|A_\beta(-\theta)\rangle
  K^\beta{}_\alpha(\theta).
\end{equation}
The action of the symmetry charges $\hat{Q}_\pm$ on the soliton
states can be obtained from the action of the quantum group
charges given in \cite{Ber91}. One finds (after a change of basis
with respect to that used in \cite{Ber91})
\begin{equation}
  \hat{Q}_\pm:|A_\alpha(\theta)\rangle\mapsto|A_\beta(\theta)\rangle
  \pi_\theta(\hat{Q}_\pm)^\beta{}_\alpha
\end{equation}
with
\begin{align}
  \pi_\theta(\hat{Q}_\pm)^+{}_+&=\hat{\epsilon}_\pm\,q^{\pm 1},&
  \pi_\theta(\hat{Q}_\pm)^-{}_-&=\hat{\epsilon}_\pm\,q^{\mp 1},\\
  \pi_\theta(\hat{Q}_\pm)^+{}_-&=c\,e^{\pm\theta/\gamma},&
  \pi_\theta(\hat{Q}_\pm)^-{}_+&=c\,e^{\mp\theta/\gamma},&
\end{align}
where $\gamma=\hb^2/(2-\hb^2)$ and
$c=\sqrt{\lambda\gamma^2(q^2-1)/2\pi i}$. We know that reflection
and symmetry transformations have to commute, which leads to the
following set of eight homogeneous linear equations for the four
entries of the reflection matrix (see \eqref{kint}):
\begin{equation}
  K^\gamma{}_\beta(\theta)\ \pi_\theta(\hat{Q}_\pm)^\beta{}_\alpha=
  \pi_{-\theta}(\hat{Q}_\pm)^\gamma{}_\beta\  K^\beta{}_\alpha,
  ~~~\forall\,\gamma,\alpha\in\{+,-\}.
\end{equation}
This set of equations is very easy to solve and one finds the
unique solution (up to an undetermined overall factor $k(\theta)$)
\begin{align}
  &K^+{}_-(\theta)=K^-{}_+(\theta)=
  \left(e^{2\theta/\gamma}-e^{-2\theta/\gamma}\right)k(\theta),\\
  &K^\pm{}_\pm=\frac{q-q^{-1}}{c}
  \left(\hat{\epsilon}_\pm\,e^{\theta/\gamma}+
  \hat{\epsilon}_\mp\,e^{-\theta/\gamma}\right)k(\theta).
\end{align}
This agrees with the soliton reflection matrix determined in
\cite{Gho94} by solving the reflection equation.

\section{Affine Toda theory\label{sectatt}}

To every affine Lie algebra $\hat{g}$ of rank $n$ there is
associated an affine Toda field theory \cite{Mik81} with Euclidean
action
\begin{equation} S = {1\over 4\pi} \int d^2z\,\partial\phi \bar{\partial}\phi
+{\lambda\over 2\pi} \int d^2z\,
\sum_{j=0}^n\exp\left(-i\hb{1\over|\alpha_j|^2} \alpha_j\cdot\phi
\right)\,,\end{equation} describing an $n$-component bosonic field
$\phi$ in two dimensions. The exponential interaction potential is
expressed in terms of the simple roots $\alpha_i$, $i=0,\dots,n$
of $\hat{g}$ (projected onto the root space of $g$). The parameter
$\lambda$ sets the mass scale and $\hb$ is the coupling constant.

For simplicity we shall restrict our attention to simply-laced
algebras, and choose the (slightly unusual) convention that
$|\alpha_i|^2=1$. With $g=a_1=su(2)$ this specializes to the
sine-Gordon model of Section \ref{sectsgm}.

\subsection{Conserved charges from conformal perturbation theory\label{sectcccpt}}

The quantum group symmetry algebra $U_q(\hat{g})$ is generated by
the topological charges
\begin{equation}
  T_j = {\hb\over 2\pi}\int_{-\infty}^\infty dx\, \alpha_j\cdot
  \partial_x\phi\,.
\end{equation}
together with the non-local conserved charges
\begin{equation}
Q_j = {1\over 4\pi c} \int_{-\infty}^\infty dx\,(J_j-H_j)\,,\qquad
\bar{Q}_j={1\over 4\pi c} \int_{-\infty}^\infty
dx\,(\bar{J}_j-\bar{H}_j)\,,\qquad j=0,1,\ldots,n,
\end{equation}
where
\begin{eqnarray}
& J_j = :\exp\left( {2i\over\hb} \alpha_j\cdot \varphi \right)\!:
\,,\qquad \bar{J}_j = :\exp\left(
{2i\over\hb} \alpha_j\cdot \bar{\varphi}\right)\!:\,,&\\[0.1in] & H_j =
\lambda{\hb^2\over \hb^2-2}:\exp\left(
i\left({2\over\hb}-\hb\right)\alpha_j\cdot \varphi
-i\hb\alpha_j\cdot \bar{\varphi} \right)\!: \,,&\\[0.1in] & \bar{H}_j =
\lambda{\hb^2\over \hb^2-2}:\exp\left(
i\left({2\over\hb}-\hb\right)\alpha_j\cdot \bar{\varphi}
-i\hb\alpha_j\cdot \varphi \right)\!:,
\end{eqnarray}
and we choose the normalization constant
$c=\sqrt{\lambda\gamma^2(q_i^2-1)/2\pi i}$  in order to obtain the
simple $q$-commutation relations given later in \eqref{qc}. The
linear combinations $\tilde{Q}_j=Q_j+\bar{Q}_j$ are
parity-invariant and thus yield conserved charges on the half-line
with Neumann boundary conditions.

We now add to the action a boundary perturbation, \begin{equation}
S_\epsilon = S_{\rm Neumann} +{\lambda_b\over 2\pi}\int dt\,
\Phi^{\rm pert}_{\rm boundary}(t)\,,\end{equation} where
\begin{equation}\Phi^{\rm pert}_{\rm boundary}(t) = \sum_{j=0}^n
\epsilon_j \exp\left( -{i\hb\over 2}\alpha_j\cdot\tilde{\phi}(0,t)
\right)\,,\end{equation} which leads to the boundary condition
\begin{equation}
\partial_x\tilde{\phi} = -i\hb\lambda_b \sum_{j=0}^n \epsilon_j
\alpha_j\exp\left( -{i\hb\over 2}\alpha_j\cdot\tilde{\phi}(0,t)
\right)\qquad {\rm at}\;\;x=0.\end{equation} By calculations
entirely analogous to those of section \ref{sectsgm} we find that,
due to this perturbation, the $\tilde{Q}_i$ are no longer
conserved, but instead satisfy \begin{equation} \partial_t
\tilde{Q}_i = {\lambda_b\epsilon_i\over 2\pi c} {\hb^2\over
\hb^2-1}\partial_t q^{T_i}\,,\end{equation} so that the new
conserved charges are
\begin{equation}\label{tcc}
  \widehat{Q}_i = Q_i + \bar{Q}_i +
  \hat{\epsilon}_i q^{T_i}\,,
\end{equation}
where, to first order
in perturbation theory,
\begin{equation} \hat{\epsilon}_i =
{\lambda_b\epsilon_i\over 2\pi c} {\hb^2\over 1-\hb^2}\,.
\end{equation}
Note that at this stage the boundary parameters $\hat{\epsilon}_i$
can still take arbitrary values. However, we shall see in
subsequent sections how the $|\hat{\epsilon}_i|$ are fixed,
leaving only a choice of signs.

The symmetry algebra of the boundary affine Toda theory generated
by the $\hat{Q}_i$, $i=0,\dots,n$, is a coideal subalgebra of
$\uqgh$ because $\Delta(\hat{Q}_i)=\hat{Q}_i\otimes
1+q^{T_i}\otimes(\hat{Q}_i-\hat{\epsilon}_i)$.

\subsection{Reflection matrices derived from quantum group
symmetry\label{sectrm}}

The conserved charges \eqref{tcc} derived in the previous section
can now be used to derive the soliton reflection matrices, as
explained in section \ref{sectsrmi}. We will illustrate this here
in the example of the vector solitons in $a_n^{(1)}$ Toda
theories. The new feature that arises which was not visible in the
sine-Gordon model is that solitons are converted into antisolitons
upon reflection off the boundary. Thus in particular the vector
solitons are reflected into solitons in the conjugate vector
representation.

Let $V^\mu_\theta$ be the space spanned by the vector solitons
with rapidity $\theta$. Choosing a suitable basis for
$V^\mu_\theta$ and defining the elementary matrices $e^{i}{}_j$ to
be the matrices with a 1 in the i-th row and the j-th column, the
representation matrices of the $\uqgh$ generators are
\begin{align}
  \pi^\mu_\theta(Q_i)&=x\,e^{i+1}{}_i,&
  \pi^\mu_\theta(\bar{Q}_i)&=x^{-1}\,e^{i}{}_{i+1},&
  \pi^\mu_\theta(T_i)&=-e^{i}{}_i+e^{i+1}{}_{i+1},
\end{align}
where $x=e^{\theta/\gamma}$ with $\gamma=\hb^2/(2-\hb^2)$ and
where we identify the indices $n+1=0$, $n+2=1$. The representation
matrices for the conjugate representation on $V^{\mub}_\theta$ are
\begin{align}
  \pi^{\mub}_\theta(Q_i)&=x\,e^{i}{}_{i+1},&
  \pi^{\mub}_\theta(\bar{Q}_i)&=x^{-1}\,e^{i+1}{}_{i},&
  \pi^{\mub}_\theta(T_i)&=e^{i}{}_i-e^{i+1}{}_{i+1},
\end{align}
The representation matrices of the symmetry generators $\hat{Q}_i$
then immediately follow from \eqref{tcc},
\begin{align}
  \pi^\mu_\theta(\hat{Q}_i)&=x\,e^{i+1}{}_i+
  x^{-1}\,e^{i}{}_{i+1}+
  \hat{\epsilon}_i\,((q^{-1}-1)\,e^{i}{}_i+(q-1)\,e^{i+1}{}_{i+1}+\one),\\
  \pi^{\mub}_\theta(\hat{Q}_i)&=x\,e^{i}{}_{i+1}+
  x^{-1}\,e^{i+1}{}_{i}+
  \hat{\epsilon}_i\,((q-1)\,e^{i}{}_i+(q^{-1}-1)\,e^{i+1}{}_{i+1}+\one),
\end{align}
where $\one$ denotes the unit matrix. Because the representation
matrices are so sparse
most of the $(n+1)^3$ components of the intertwining equation
\eqref{kint} for the $K$-matrix are trivial, leaving only the
$2n(n+1)$ equations
\begin{align}
  0&=\hat{\epsilon}_i(q^{-1}-q)K^i{}_i+x\,K^i{}_{i+1}-x^{-1}\,K^{i+1}{}_i,
  \label{s3}\\
  0&=K^{i+1}{}_{i+1}-K^i{}_i,\label{s4}\\
  0&=\hat{\epsilon}_i\,q\,K^i{}_j+x^{-1}\,K^{i+1}{}_j,~~~j\neq
  i,i+1,\label{s1}\\
  0&=\hat{\epsilon}_i\,q^{-1}\,K^j{}_i+x\,K^j{}_{i+1},~~~j\neq
  i,i+1.\label{s2}
\end{align}
The equations in \eqref{s1} and \eqref{s2} can be used to
determine all upper triangular entries in terms of $K^1{}_2$ and
all lower triangular entries in terms of $K^2{}_1$. Then the
equations \eqref{s3} determine the diagonal entries. Finally the
fact that all diagonal entries must be the same according to
\eqref{s4} not only determines $K^2{}_1$ in terms of $K^1{}_2$ but
also requires that either $\hat{\epsilon}_i=0$ for all $i$ or that
$|\hat{\epsilon}_i|=1$ for all $i$. If all $\hat{\epsilon}_i=0$
then $K$ can be an arbitrary diagonal matrix. If all
$|\hat{\epsilon}_i|=1$ then one obtains the non-diagonal solution
\begin{align}\label{k}
  K^i{}_i(\theta)&=\left(
  q^{-1}\,(-q\,x)^{(n+1)/2}-
  \hat{\epsilon}\,q\,(-q\,x)^{-(n+1)/2}
  \right)
  \frac{k(\theta)}{q^{-1}-q},\notag\\
  K^i{}_j(\theta)&=\hat{\epsilon}_i\cdots\hat{\epsilon}_{j-1}\,
  (-q\,x)^{i-j+(n+1)/2}\,k(\theta),~~~~~~\text{for }j>i,\\
  K^j{}_i(\theta)&=\hat{\epsilon}_i\cdots\hat{\epsilon}_{j-1}\hat{\epsilon}\,
  (-q\,x)^{j-i-(n+1)/2}\,k(\theta),~~~~~\text{for }j>i,\notag
\end{align}
which is unique up to the overall numerical factor $k(\theta)$. We
have defined
$\hat{\epsilon}=\hat{\epsilon}_0\hat{\epsilon}_1\cdots\hat{\epsilon}_n$.
Note that this agrees with the solution found by Gandenberger in
\cite{Gan99b}.

It is interesting to note that the restriction on the possible
values of the boundary parameters $\hat{\epsilon}_i$ which we
found above agrees with the restrictions which were found in
\cite{Bow95} from the requirement of classical integrability.

The reflection matrices for solitons in the multiplets
corresponding to the other fundamental representations could be
determined either in the same manner as above or by a fusion
procedure. Furthermore there will be boundary bound states which
may  transform non-trivially under the quantum group symmetry and
again their reflection matrices can be obtained by using the
intertwining property or by boundary fusion. We refer the reader
to \cite{Del01} where similar calculations have been done for the
rational reflection matrices which have a Yangian symmetry.

\subsection{Construction of conserved charges from reflection
matrices\label{sectccrm}}

In section \ref{sectcccpt} we derived expressions for the symmetry
generators $\hat{Q}_i$ by using first-order boundary conformal
perturbation theory. These symmetry generators allowed us to
determine the reflection matrices for the vector solitons in
$a_n^{(1)}$ Toda theory in section \ref{sectrm}. This success can
be taken as confirmation of the correctness of the expressions
\eqref{tcc} for the symmetry generators. However one might worry
that higher-order perturbation theory might produce additional
terms which, while not visible in the vector representation, might
be required to guarantee the commutation of the generators with
the reflection matrices in higher representations. In order to
rule this out, we will now rederive the expressions for the
generators $\hat{Q}_i$ using the construction introduced in
section \ref{sectcsa}. We will show how to extract the $\hat{Q}_i$
from the $B^\mu_\theta$ by expanding to first order in
$x=\exp(\theta/\gamma)$.

The construction requires the L-matrices that are obtained from
the universal R-matrix according to \eqref{l}. Luckily we will not
need to work with the rather involved expression for the full
universal R-matrix. Instead we will introduce the spectral
parameter $x$ and expand to first order in $x$. For this purpose
let us introduce the homomorphism
$\Psi_x:\uqgh\rightarrow\uqgh[x,x^{-1}]$, defined by
\begin{align}
  \Psi_x(Q_i)&=x\,Q_i,&\Psi_x(\bar{Q}_i)&=x^{-1}\,\bar{Q}_i,&
  \Psi_x(T_i)&=T_i.
\end{align}
This will be useful later because
\begin{equation}
  \pi^\mu_\theta=\pi^\mu_0\circ\Psi_x~~~\text{for
  }~~x=e^{\theta/\gamma}.
\end{equation}
We introduce the spectral parameter dependent universal R-matrix
as
\begin{equation}
  \ur(x)=(\Psi_x\otimes\id)(\ur).
\end{equation}
It satisfies
\begin{equation}\label{urp}
  \ur(x)\left((\Psi_x\otimes\id)\circ\Delta\right)(Q)=
  \left((\Psi_x\otimes\id)\circ\Delta^{\text{op}}\right)(Q)
  \ur(x)~~~~\forall Q\in\uqgh.
\end{equation}
We want to use this property to determine the expression for
$\ur(x)$ to linear order in $x$. For this purpose we need the
relations between the $\uqgh$ generators:
\begin{align}
  &[T_i,Q_j]=\ab_i\cdot\ab_j\,Q_j,~~~~~~
  [T_i,\bar{Q}_j]=-\ab_i\cdot\ab_j\,\bar{Q}_j\notag\\
  &Q_i\bar{Q}_j-q^{-\alpha_i\cdot\alpha_j}\bar{Q}_j
  Q_i=\delta_{ij}\,\frac{q^{2 T_i}-1}{q_i^{2}-1},\label{qc}
\end{align}
where $q_i=q^{\ab_i\cdot\ab_i/2}$. The generators also satisfy
Serre relations which we will not need however. The coproduct
$\Delta$ is defined by
\begin{align}\label{delta}
  \Delta(Q_i)&=Q_i\otimes 1+q^{T_i}\otimes Q_i,\notag\\
  \Delta(\Qb_i)&=\Qb_i\otimes 1+q^{T_i}\otimes \Qb_i,\\
  \Delta(T_i)&=T_i\otimes 1+1\otimes T_i.\notag
\end{align}
Using this information we find that
\begin{equation}
  \ur(x)=\left(1\otimes 1+x\,\sum_{l=0}^n(1-q_l^2)Q_l\otimes\bar{Q}_l
  \,q^{-T_l}\right)
  q^{\sum_{j,k=1}^n g_{jk}T_j\otimes T_k}+\mathcal{O}(x^2),
\end{equation}
or, equivalently,
\begin{equation}
  \ur(x)=q^{\sum_{j,k=1}^n g_{jk}T_j\otimes T_k}
  \left(1\otimes 1+x\,\sum_{l=0}^n(1-q_l^2)\,q^{-T_l}\,Q_l
  \otimes\bar{Q}_l\right)
  +\mathcal{O}(x^2),
\end{equation}
where $g_{jk}\ \alpha_k\cdot\alpha_l=-\delta_{jl}$.

We now specialize to the vector representation of $a_n^{(1)}$ and
find the $L$-operators according to eqns. \eqref{l},
\begin{align}
  L^\mu_\theta&=(\pi^\mu_\theta\otimes\id)(\ur)=(\pi^\mu_0\otimes\id)(\ur(x))\\
  &=q^A\left(\one\otimes
  1+x\,\sum_{l=0}^n(q^{-1}-q)\,e^{l+1}{}_l\otimes\bar{Q}_l\right)
  +{\mathcal O}(x^2),\\
  \bar{L}^{\mub}_\theta&=(\pi^{\mub}_{-\theta}\otimes\id)(\ur^{\text{op}})
  =(\pi^{\mub}_0\otimes\id)(\ur(x)^{\text{op}})\\
  &=\left(\one\otimes
  1+x\,\sum_{l=0}^n(q^{-1}-q)\,e^{l+1}{}_l\otimes{Q}_l\right)q^{-A}
  +{\mathcal O}(x^2),
\end{align}
where $A=\sum g_{jk}(-e^j{}_j+e^{j+1}{}_{j+1})\otimes T_k$. We
also expand the reflection matrix $K^\mu(\theta)$ given in
\eqref{k} to first order in $x$,
\begin{equation}
  K^\mu(\theta)=B\left(\one+x\,\sum_{l=0}^n\hat{\epsilon}_l\,
  (q^{-1}-q)\,e^{l+1}{}_l\right)+{\mathcal O}(x^2),
\end{equation}
where $B=-\hat{\epsilon}\,q\,(-q\,x)^{-(n+1)/2}/(q-q^{-1})$.
Putting it all together according to eq.\eqref{b} gives
\begin{equation}
  B^\mu_\theta=B+x\,\sum_{l=0}^n(q^{-1}-q)\,e^{l+1}{}_l\otimes
  \left(Q_l+\bar{Q}_l+\hat{\epsilon}_l\,q^{T_l}\right)+{\mathcal O}(x^2).
\end{equation}
Thus we can read off our generators $\hat{Q}_i$ from the
non-vanishing entries of the matrix $B^\mu_\theta$ at first order
in $x$. This proves that their action does indeed commute with all
reflection matrices because we had shown this for the entries of
the matrix $B^\mu_\theta$ already in section \ref{sectcsa}. They
thus generate a symmetry algebra of $a_n^{(1)}$ affine Toda theory
on the half-line.

\section{Discussion\label{sectdis}}

In this paper we have derived non-local conserved charges for the
sine-Gordon model and $a_n^{(1)}$ affine Toda field theories on
the half-line and have shown how to use these to determine the
soliton reflection matrices by solving the linear intertwining
equations.

The calculations in conformal boundary perturbation theory used to
derive the non-local charges in sections \ref{sectqgcgbc} and
\ref{sectcccpt} may be of interest in themselves because of the
way in which the perturbation theory for the model on the
half-line is embedded into that for the model on the whole line.

The calculations should easily generalize to the Toda theories for
arbitrary affine Kac-Moody algebras $\hat{g}$, allowing us to
derive the hitherto unknown soliton reflection matrices in these
theories. This has recently been carried through for the case of
$\hat{g}=d_n^{(1)}$ in \cite{A}. One can then derive also the
particle reflection amplitudes, as was done for the $a_n^{(1)}$
Toda particles in \cite{Del99}.

The reconstruction of the symmetry algebra from the reflection
matrix described in section \ref{sectcsa} is of wider
applicability. For example we can apply it to the diagonal
reflection matrices $K^\mu_\theta:V^\mu_\theta\rightarrow
V^\mu_\theta$ found in \cite{dev93}. This gives the symmetry
algebra of the $su(n)$ spin chain on the half-line and is the
subject of forthcoming work with Phil Isaac. We expect that there
will also be new kinds of boundary conditions in affine Toda field
theory which preserve this symmetry algebra, generalizing the
boundary condition found in \cite{Del98b}. It will be interesting
to find these, in particular as some of the corresponding soliton
reflection matrices have already been calculated in the continuum
limit of the $su(n)$ spin chain \cite{Doi99}.

We have demonstrated that one can obtain solutions of the
reflection equation by solving intertwining equations for
representations of suitable coideal subalgebras of $\uqgh$. This
is of great practical importance because the linear intertwining
equations are much easier to solve than the reflection equation
itself.

The interesting mathematical problem therefore now presents itself
of classifying all relevant coideal subalgebras of $\uqgh$ and
their representations. We expect this to lead to a classification
of all trigonometric reflection matrices in analogy to the
classification of trigonometric $R$-matrices in terms of
representations of quantum affine algebras. The required
properties of the coideal subalgebras are that they should be
``small enough'' so that the intertwiners exist, but also ``large
enough'' so that the tensor product representations are
generically irreducible.

In the rational case, generators for the relevant coideal
subalgebras of the Yangians $Y(g)$ have been constructed in
\cite{Del01}. In that case one has to consider the involutive
automorphisms $\sigma$ of the Lie algebra $g$ which lead to
symmetric pairs $(g,g^\sigma)$. The coideal subalgebra of $Y(g)$
is then a quantization of the corresponding twisted polynomial
algebra. We shall denote these algebras $Y(g,g^\sigma)$ and refer
to them as twisted Yangians. Twisted Yangians for $g=su(n)$ have
already been described in \cite{Mol96} for $g^\sigma=so(n)$ and
$g^\sigma=sp(n)$ and in \cite{Min99} for $g^\sigma=su(m)\oplus
su(n-m)\oplus u(1)$.

One might therefore hope in the trigonometric case to arrive at a
theory of twisted quantized affine algebras
$U_q(\hat{g},\hat{g}^\sigma)$. In the classical case twisting by
an inner automorphism leads to isomorphic algebras, leaving only
the known twisted affine algebras based on Dynkin diagram
automorphisms. In the quantum case however, where a particular
Cartan subalgebra is singled out by the quantization, new algebras
arise.

In the non-affine case the analogous construction of coideal
subalgebras of $U_q(g)$ from involutions has been studied in
\cite{Let01}. The motivation in this case is that they lead to
quantum symmetric pairs and thus to quantum symmetric spaces and
their associated $q$-orthogonal polynomials. These algebras
include those constructed in \cite{Nou} by reflection matrix
techniques closer to our construction in section \ref{sectcsa}. We
will be seeking affine generalizations of these works.

Preliminary results from this paper were presented at the 5th
Workshop on CFT and integrable models in Bologna in September 2001
and at seminars in Tokyo, Sydney, Brisbane and York in October and
November 2001.

\noindent{\bf Acknowledgements:}

Part of this work was performed during a short stay of GWD at the
University of Tokyo supported by the JSPS and the Royal Society.
GWD would like to thank Alex Molev, Ruibin Zhang, Mark Gould, and
Yao-Zhong Zhang for their hospitality in Sydney and Brisbane. GWD
is supported by an EPSRC advanced fellowship.

\end{document}